\documentclass[authoryear]{elsarticle}\usepackage[]{graphicx}\usepackage[]{color}
\makeatletter
\def\maxwidth{ %
  \ifdim\Gin@nat@width>\linewidth
    \linewidth
  \else
    \Gin@nat@width
  \fi
}
\makeatother

\definecolor{fgcolor}{rgb}{0.345, 0.345, 0.345}

\usepackage{framed}
\makeatletter
 {\par\unskip\endMakeFramed%
 \at@end@of@kframe}
\makeatother

\definecolor{shadecolor}{rgb}{.97, .97, .97}
\definecolor{messagecolor}{rgb}{0, 0, 0}
\definecolor{warningcolor}{rgb}{1, 0, 1}
\definecolor{errorcolor}{rgb}{1, 0, 0}
\newenvironment{knitrout}{}{} 

\usepackage{alltt}

\usepackage{array}
\usepackage[colorinlistoftodos]{todonotes}
\usepackage[sc]{mathpazo} 
\usepackage[hang, small,labelfont=bf,up,textfont=it,up]{caption} 
\usepackage{booktabs} 
\usepackage{float} 
\usepackage{paralist} 
\usepackage{graphicx}
\usepackage{bm}
\IfFileExists{upquote.sty}{\usepackage{upquote}}{}
\begin{document}
\begin{frontmatter}
\title{Spectral goodness of fit for network models} 

\author[bu]{Jesse Shore\corref{cor1}}
\author[bu]{Benjamin Lubin}

\cortext[cor1]{jccs@bu.edu}
\address[bu]{Boston University School of Management}

\begin{abstract}
\noindent  
We introduce a new statistic, 'spectral goodness of fit' (SGOF) to measure how well a network model explains the structure of an observed network. SGOF provides an absolute measure of fit, analogous to the standard $R^2$ in linear regression.  Additionally, as it takes advantage of the properties of the spectrum of the graph Laplacian, it is suitable for comparing network models of diverse functional forms, including both fitted statistical models and algorithmic generative models of networks.  After introducing, defining, and providing guidance for interpreting SGOF, we illustrate the properties of the statistic with a number of examples and comparisons to existing techniques.  We show that such a spectral approach to assessing model fit fills gaps left by earlier methods and can be widely applied.  


\end{abstract}

\date{\today}
\end{frontmatter}



\section{Introduction}
Models of network structure play several important roles in contemporary science.  Parametric statistical models of network structure and dynamics allow inferences to be made about dependencies among network ties, network position, and nodal and dyadic covariates \citep{frank1986markov,anderson1992building, snijders2001statistical, schweinberger2003settings,handcock2003assessing,doreian2005generalized, hunter2006inference,steglich2010dynamic}.   Algorithmic generative models illustrate how complex macroscopic structure can arise from simple and often local rules \citep{watts1998collective,vazquez2003growing,saramaki2004scale}.  Despite the importance and diversity of research within both the model based inference and generative algorithms categories, one aspect of network model-based research that has been relatively slow to develop is that of assessing goodness of fit, or how well a given model describes the empirical data being modeled. Moreover, the methods that are commonly used to assess fit within one type of model may be uncommon or unavailable in another, making it difficult to integrate research techniques and results across scholarly communities.

The purpose of this paper is therefore to define a new measure of goodness of fit that substantially fills the gaps left by current methods.   In particular, leveraging the features of the spectrum of the graph Laplacian, we define a new goodness of fit statistic that measures the percent improvement a network model makes over a null model in explaining the structure in the observed data. As such, we provide a goodness of fit measure that can be applied across modeling techniques and which provides an absolute measure of goodness of fit for the model to the observed network data.

\subsection{Existing methods}
Commonly used existing methods for assessing goodness of fit can be roughly classified into two groups: one based on comparing structural statistics from networks simulated from a fitted model to structural statistics from the observed network \citep{hunter2008goodness,schweinberger2012statistical}, and the other based on a model's likelihood function, exemplified by the Akaike Information Criterion \citep{hunter2008goodness}.  
\subsubsection{Structural-statistics comparisons}
The most commonly used method of assessing goodness of fit (GOF) is the structural statistics approach, which  is implemented in software for estimating Exponential Random Graph Models (ERGMs) as well as dynamic actor-oriented models (also known as 'Siena' models).  Although not done in a hypothesis testing framework, important algorithmic models \citep[e.g.][]{watts1998collective} have also been described in terms of how well the algorithm reproduces the subgraph statistics in observed networks.

In this approach, after fitting a model, it is necessary to generate a large number of simulated networks based on that model.  At that point comparisons can be made between the observed and the simulated networks.  The modeler might ask if the observed number of closed triads (or distribution of closed triads over the nodes) could have been drawn from the distribution defined by the simulated networks, or if the observed degree distribution could have been drawn from the distribution of degree distributions in the simulated networks, or any number of other questions of fit between statistics describing the observed and simulated networks.  If the structures in the observed network are very unlikely to have been generated by the fitted model, the modeler can reject the hypothesis that the model fits well.  

The subgraph-statistical approach has many advantages.  By specifying different structural statistics to compare, the approach can be readily adapted to different specific questions of model fit.  For example, one researcher may have a theoretical reason to emphasize the length of geodesics, while another may focus on triadic closure.  The results of such an analysis are also easy to interpret and lend themselves to graphical representation and inspection (as in \cite{hunter2008goodness}).

On the other hand, this method also has limitations. Even if the theoretical focus of a given researcher is on a single structural issue, say, modeling geodesics, the \textit{overall} fit of the model to the whole network is still important.  A model that accurately reproduces the distribution of geodesics but does not reproduce the overall structure of the network is probably inferior to one that captures the geodesic distribution and the overall structure simultaneously.  

The difficulty in the subgraph-statistical approach is that it is not clear how to measure the overall structure of the network, except in terms of a list of its statistics.  This approach necessarily decomposes the goodness of fit of a whole model into multiple goodness of fit tests on specific features of the model. Theoretically, this is problematic; practically, the validity of the goodness of fit assessment depends heavily on which statistics are specified by the researcher for examination.  In a sense, in order to construct a valid goodness of fit test, the researcher is required to know a priori what the important statistics are for a given observed network; this is sometimes a nonsensical requirement, as goodness of fit tests are often undertaken exactly because the research does \textit{not} know whether a given set of statistics (those described by the model parameters) are a good descritption of a network.  The pragmatic solution is to use a commonly accepted set of statistics (\cite{hunter2008goodness} provides a good argument for one such set), but the possibility remains that important aspects of structure are not considered in such a goodness of fit test.  

Additionally, assessing model fit in terms of subgraph statistics does not provide a means of selecting between two models that are both rejected or both not rejected: it provides neither a relative nor an absolute measure of fit by which such a comparison could be made.  Finally, it is difficult to compare published results from different studies when they do not report the same subgraph statistical tests or analysis.  

\subsubsection{Akaike Information Criterion}
Likelihood-based approaches, exemplified by the Akaike Information Criterion (AIC) (available for example, to users of the \texttt{ergm} package in \texttt{R} \citep{ergmMan, ergmArt}), fills some of the gaps left by hypothesis tests on structural statistics.  The AIC is a well-known tool for model choice based that provides a relative measure of goodness of fit.  

There are several limitations of the AIC as well.  First, many models do not have a well-defined AIC, including ERGMs that are conditioned on having the exact number of edges present in the observed network, as well as models of networks that were not estimated from a statistical model at all (cases that we consider in more detail below). 

Second, the AIC measures goodness of fit of all model parameters to all data, which may not always be what is desired.  There are sometimes cases when a researcher wants to know if some model could have generated the observed pattern of ties alone, rather than whether the model could have jointly generated the ties and nodal and dyadic covariates.  To briefly cite an example we discuss below, in specifying a model with a homophily parameter, the researcher may want to know how well the model explains the pattern of ties, rather than how well the model describes the homophily.  AIC provides information on the latter, but not the former.

Third, like the structural-statistics approach to which it is related, one cannot know if there are omitted variables that would have improved the fit of the model.  While the AIC can compare the relative quality of two models in certain senses, it cannot say whether either model is any good in in an absolute sense. 


\subsection{Spectral Goodness of Fit}

Given the tools already available to network modelers, a desirable measure of goodness of fit would have the following properties:
\begin{itemize}
\item it would provide an absolute (not relative) measure of goodness of fit 
\item it would not require the modeler to know the true model or which structural statistics are important in the observed network
\item It would allow comparison of a wide range of models, including those without likelihood functions or even without statistical parameterizations 
\end{itemize}

In other words, it would have properties analogous to the $R^2$ used in standard linear regression. Here, we propose such a statistic: spectral goodness of fit (SGOF).

Throughout the rest of this article we make several assumptions.  We consider only undirected networks explicitly, although we discuss extensions to directed networks in the final section, below. Additionally, in proposing to assess goodness of fit, we assume that a researcher has data on an observed network and has fit (or otherwise chosen) a model of network structure to that data. We do not make any assumptions about the functional form of that model or even whether the model is parametric at all, but we do assume that the researcher can generate simulated networks based on the fitted model. 
\subsection{Computer Code}
We have made computer code for calculating SGOF and visualizing the results of the analysis available as an R package, \texttt{spectralGOF}\footnote{Available at http://people.bu.edu/jccs}.


\section{The spectrum of the graph Laplacian}
\subsection{Definitions and notation}
Networks are frequently represented as square adjacency matrices (which we will denote by $A$), such that if there is a link from node $u$ to node $v$, then $A_{uv}>0$.  For the purposes of this article, we are considering only undirected networks, so $A_{uv}=A_{vu},  \forall u \forall v$.

The Laplacian matrix is a transformation of the adjacency matrix given by $L = D - A$, where $D$ is the 'degree matrix,' containing the row sums of $A$ on its diagonal and zeros elsewhere.  The spectrum of $L$ is the ordered multiset of eigenvalues, $\boldsymbol{\lambda}$, such that $0=\lambda_1 \leq \lambda_2 \dots \leq \lambda_n $.  There is one Laplacian eigenvalue (hereafter, for brevity, 'eigenvalues' and 'spectrum' always refer to the eigenvalues of the Laplacian) equal to zero for every connected component in the network \citep{brouwer2011spectra}.  Therefore, $\lambda_1$ is always $0$.

The sum of all eigenvalues is equal to the total weight of all edges in the network: 
\begin{equation}\label{sumOfEvs}\sum_{i=1}^n \lambda_i = \sum_{u=1, v=1}^n{A_{uv}}\end{equation}

\subsection{The spectrum of the Lapacian as a representation of network structure}
The spectrum is a ``graph invariant,'' meaning that if two networks are isomorphic\footnote{Isomorphic networks have the same structure. They could be represented by the same adjacency matrix after permuting the rows and columns and disregarding any ``labels'' or names of the nodes.}, then they have the same spectrum.   
The spectrum is also a compact representation of a great deal of structural information, and spectral techniques (sometimes including analysis of both the spectrum and its associated eigenvectors) have thus been used extensively to characterize the structure of complex networks \citep{pothen1990partitioning,newman2006modularity} and to compare and recognize complex objects in other applications such as facial recognition in computer vision \citep{turk1991eigenfaces,belkin2003Laplacian}. The properties of the Laplacian spectrum have been studied extensively \citep[see][for relatively accessible mathematical overviews]{mohar1991Laplacian,brouwer2011spectra, chung1997spectral} and a full treatment is well beyond the scope of this article.  However, to provide context for our definition of the spectral goodness of fit statistic, we do provide some basic intuition for the connection beween the spectrum and network structure in the following paragraphs.

As we have already noted, the number of components is reflected in the spectrum by the number of zeros.  The magnitude of the smallest non-zero eigenvalue is related to the minimum number of ties (how much total weight) that would have to be cut (that is, removed from the network) to divide the network into two disconnected components and is known as the ``algebraic connectivity'' of a network \citep{fiedler1973algebraic}. The magnitudes of the next smallest eigenvalues represent the relative modularity of the next-most macroscopic community structure of a network. \cite{donetti2006optimal} illustrate this logic as follows. Imagine a network comprising four totally disconnected components.  Its spectrum would contain four eigenvalues equal to zero.  If we perturb this network by connecting the components with a small number of ties \citep{cvetkovic1997eigenspaces}, such that they are no longer disconnected, then rather than having one eigenvalue equal to zero for each component, we would have one \textit{small} eigenvalue for each modular cluster \citep{donetti2006optimal}, one of which would be zero (as there would be one component, and thus one eigenvalue equal to zero).  The more weight that was added between the components, the larger the eigenvalues would become.

The sizes of successively larger eigenvalues provide information on successively finer divisions of the network into smaller sub-communities. In general, a common interpretation of the magnitudes of eigenvalues of the Laplacian is one of correspondence to the relative weight removed by a series of minimum cuts of the network \citep[for a more detail, see, e.g.][]{bollobas2004graphs}.   The largest eigenvalue therefore contains information about the number of ties incident to the single most highly connected node \citep{schur1923uber,brouwer2011spectra}.

\subsection{Normalizing the spectrum}
The shape of the spectrum describes how the total tie strength in a given network is structured relative to other networks with the same total amount of tie strength (density).  Given this, in the definition of the spectral goodness of fit (SGOF) statistic below, we normalize all spectra to sum to unity. 

As equation \ref{sumOfEvs} indicates, the sizes of the eigenvalues are sensitive to the density of the network.  More specifically, given an adjacency matrix, $A$, let us  denote by $\hat{A}$ a normalized version of $A$. 
\begin{equation}
\hat{A}= \frac{A}{\sum{A}}
\end{equation}

Likewise, as $\boldsymbol\lambda$ is the vector of eigenvalues of $A$, let $\hat{\boldsymbol\lambda}$ denote the vector of eigenvalues of $\hat{A}$, which can also be calculated by normalizing  $\boldsymbol\lambda$.

\begin{equation}
\hat{\boldsymbol\lambda}= \frac{\boldsymbol\lambda}{\sum{\boldsymbol\lambda}}
\end{equation}

An increase in the density of $A$ that does not result in changes to $\hat{A}$ (i.e., multiplying all entries in $A$ by a non-zero scalar constant) also does not change $\hat{\boldsymbol\lambda}$.  In other words, such a change only alters the size and not the shape of the spectrum. On the other hand, an increase in the density of $A$ that does result in changes to  $\hat{A}$ (i.e., adding new ties or increasing the strength of certain ties and not others) both increases the sizes of $\boldsymbol\lambda$ and changes its shape: it results in a changed $\hat{\boldsymbol\lambda}$ as well.


\section{Spectral Goodness of Fit}

\subsection{Spectral distance}
Given the structural information contained in the spectrum, the Euclidean distance between two spectra is frequently used as a measure of the structural similarity of two matrices \citep{cvetkovic2012spectral}.  The Euclidean spectral distance ($ESD$) can be written as 
$||\hat{\boldsymbol\lambda}^{A}-\hat{\boldsymbol\lambda}^{B}||$, where the normalized full spectra of graphs $A$ and $B$ are given by $\hat{\boldsymbol\lambda}^{A}$ and $\hat{\boldsymbol\lambda}^{B}$, and the double bars denote the the vector norm.

\newcommand{\nsim}{\ensuremath{N_{sim}}}
We wish to apply this notion of distance to our network models, but such models do not themselves have spectra.  However, if networks can be simulated from or otherwise generated by the model, spectra for these networks can be calculated.  It is the distance between these spectra and the observed spectrum that we will consider.  If we have, say, $\nsim=1000$ simulated networks, we can calculate the mean spectral distance between the simulated networks and the observed network, as well as other distributional statistics, such as the $5^{th}$ and $95^{th}$ percentiles of the spectral distance between simulations and the observed network.


Formally, after normalizing the spectra as above, let us call the absolute value of the difference between the $i^{th}$ observed eigenvalue and the $i^{th}$ eigenvalue from the $k^{th}$ simulated network an 'error.'  
\begin{equation}
\epsilon_i = \left|\hat{\lambda}_i^{obs}-\hat{\lambda}_i^{sim_k}\right| 
\end{equation}
In this context then, $ESD$ is the square root of the sum of squared errors. 
\begin{equation}
\label{singleError}
ESD_{obs,sim_k}=\left|\left|\hat{\boldsymbol\lambda}^{obs}-\hat{\boldsymbol\lambda}^{sim_k}\right|\right|=\sqrt{\sum_i{(\epsilon_i)^2}}
\end{equation}
The mean Euclidean spectral distance, $\overline{ESD}$,  is then defined as arithmetic mean of the ESDs from each of the individual simulated networks.
\begin{equation} \overline{ESD}_{obs,sim}=\frac{1}{\nsim} \sum_{k=1}^{\nsim}ESD_{obs,sim_k}\end{equation}

\subsection{Definition of null model}
For network models we propose that goodness of fit be measured as an improvement in fit relative to a naive null model.  It is therefore necessary to calculate the errors under the naive model and the fitted model for some number of simulated networks.

The natural null model for dichotomous networks is the density-only model, also known as the Bernoulli model or Erd\H{o}s-R\'enyi model, simulatations from which are random networks with the same expected density as the observed network.  For the remainder of this article, we adopt the density-only model as a null model, but we note that any other model could be substituted in its place. 

One situation where the Erd\H{o}s-R\'enyi model would not be appropriate as a null model is the case where the measurement of the observed network was by means of a survey instrument that specified the number of alters each respondant was to nominate ('name five people you discuss important matters with'). In this case a degree-regular random graph (one in which each node has the same degree) would be the appropriate null model.  Likewise, if the observed data is weighted, the null model should also be weighted. In general, the null model should be the maximum entropy model generating networks in the same class as the observed data.

\subsection{Definition of SGOF}
To calculate the Spectral Goodness of Fit ($SGOF$), we simply divide the mean Euclidean spectral distance under the fitted model by the mean Euclidean spectral distance under the null model, and subtract the result from one.

\begin{equation}
\label{sgofdef}
SGOF=1- \frac{ \overline{ESD}_{obs,fitted}}{\overline{ESD}_{obs,null}} 
\end{equation}

Additionally, given that models of networks imply a probability distribution of networks generated from the model, it is advisable to report SGOF calculated using the $5^{th}$ and $95^{th}$ percentile results for $ESD$ under the fitted model.  Below, we report these in parentheses after the SGOF calculated using the mean as in equation \ref{sgofdef}.  This confidence interval provides an indication of the dispersion of goodness of fit inherent in a fitted model.

Although the mean SGOF of the null model is defined to be zero, it is advisable to report the $5^{th}$ and $95^{th}$ percentile results for the null model as well.  The reason for this is that the width of this 90\% confidence interval provides useful information for interpreting the SGOF of fitted models.  If an observed network is not highly structured, the 90\% confidence interval for the null model's SGOF will be very wide, extending, say, from $-0.5$ to $0.5$, reflecting the fact that the observed network is not far from random.  For observed networks with a great deal of structure, the 90\% confidence interval for the null model's SGOF will be narrow, extending for example only from $-0.001$ to $0.001$.

\subsection{Interpretion of SGOF}
The SGOF measures the amount of observed structure explained by a fitted model, expressed as a percent improvement over a null model, where structure means deviation from randomness.  The observed spectrum will be distant from the spectrum of the null model in as much as the observed network has structure that is non-random.  The SGOF is thus a summary measure of the percent of the observed structure that is explained by the fitted model. 

\subsubsection{Bounds for SGOF}

Like $R^2$, SGOF is bounded above by one, when the fitted model exactly describes the structural data. Likewise, an SGOF of zero means no improvement over the null model.  Finally, as with $R^2$, SGOF can be unboundedly negative\footnote{In normal practice, however, the fitted model for $R^2$ is an ordinary least squares linear regression with a free intercept parameter; in this typical case, $R^2$ is bounded below by zero.} if the spectrum of the fitted model is more distant from the observed spectrum than is the spectrum of the null model.   If the SGOF is negative, it is therefore evidence that the null model (an Erd\H{o}s-R\'enyi random graph) is a better approximation of the observed network than the fitted model under consideration.
This is likely to occur in cases where the observed network is not highly structured (and thus very similar to the null model), and the fitted model is (incorrectly) highly structured.  If the observed network is not structured, then while $\overline{ESD}_{obs, fitted} > 0$, $\overline{ESD}_{null} \rightarrow 0$ and by equation \ref{sgofdef}, $SGOF \rightarrow -\infty$.   For ordinary cases involving an observed network that contains structure to be explained and sensible model specifications, however, SGOF will fall between zero and one.

\section{Applications and comparisons to existing methods}

In this section, we illustrate the spectral goodness of fit method with several examples chosen to highlight its strengths and weaknesses with respect to existing methods.  

\subsection{Comparison with structural statistics: e.coli}
It is frequently the case that a researcher does not ever discover the 'true' model underlying the formation of an observed network, but rather is only able to approximate the truth with several theoretically plausible candidate models.  In such cases it is useful to have quantitative evidence about model goodness of fit to help adjudicate the decision.  Structural statistical tests can sometimes play this role, but as mentioned above, it may also be the case that all models under consideration are rejected (or supported) by the test, and more information is therefore needed.

This example considers such a situation by comparing two specifications of a model of the degree distribution of the \textit{e. coli} genetic regulatory network \citep{shen2002network}, both in the ERGM framework.  

\begin{table*}[htbp]
\caption{Comparison of Spectral Goodness of Fit to structural hypothesis testing for the \textit{e. coli} genetic regulatory network}
\label{ECtab}
  \begin{tabular}{m{1.5in} r m{.4in} m{1.2in}}
  \hline\hline

Observed Network& & &
\includegraphics[width=1.1in,height=1.1in]{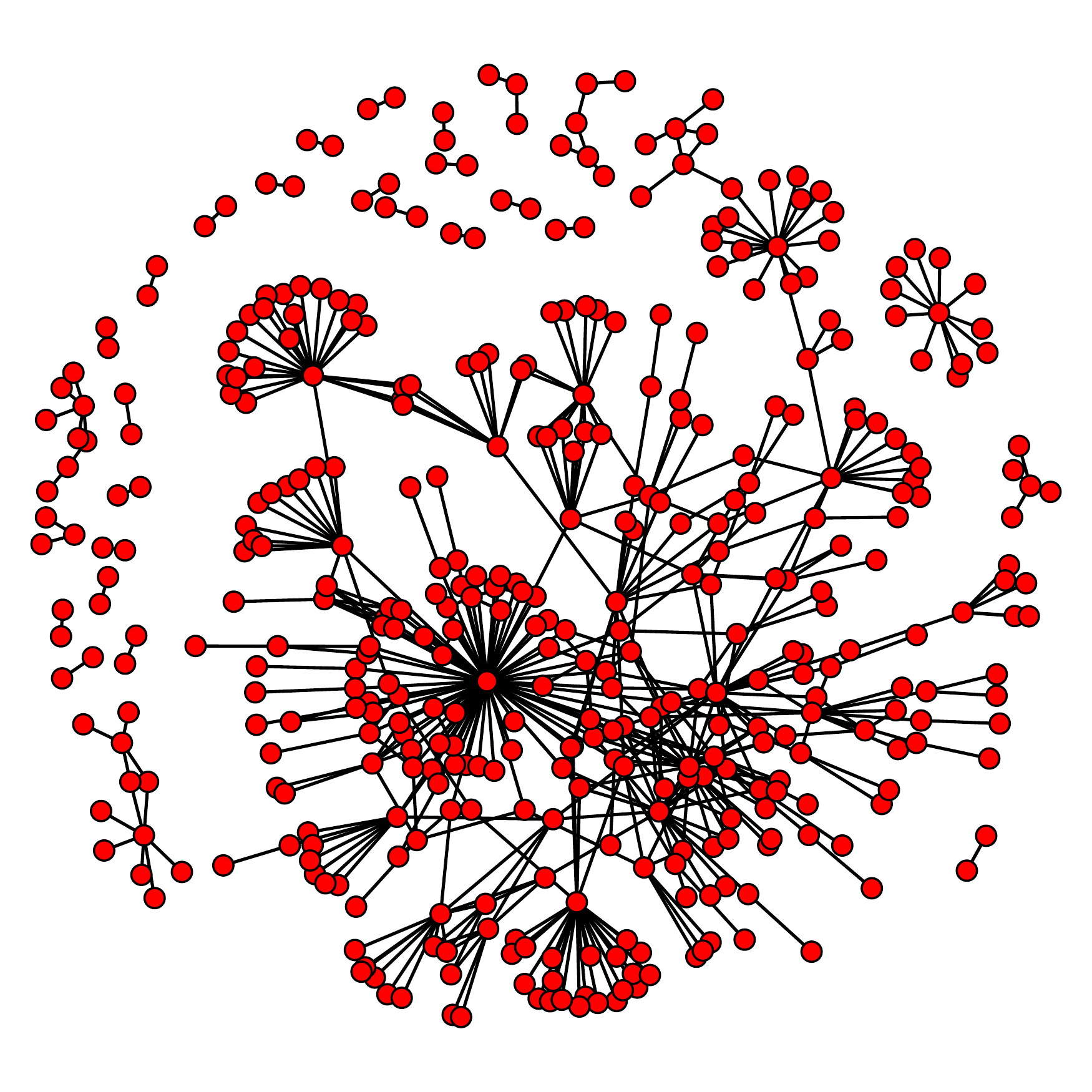} 
\\\hline\\     & SGOF     & Struc. h-test  &  Simulated Network\\\cline{2-4}Null model& 0 (-0.02, 0.025) & reject &
\includegraphics[width=1.1in,height=1.1in]{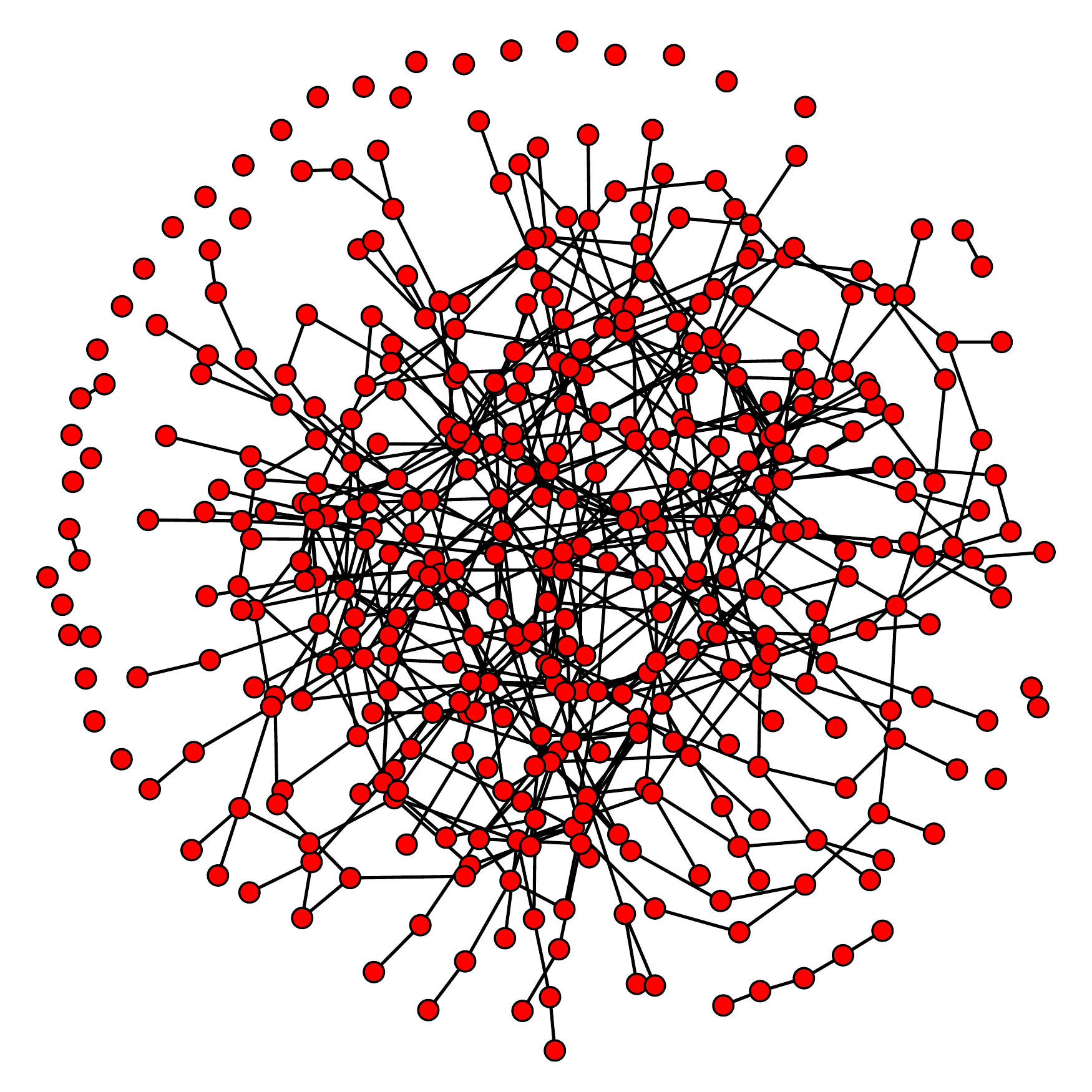} 
\\Geom. weighted degree  (curved exponential family) & 0.242 (0.167, 0.33)  & reject &
\includegraphics[width=1.1in,height=1.1in]{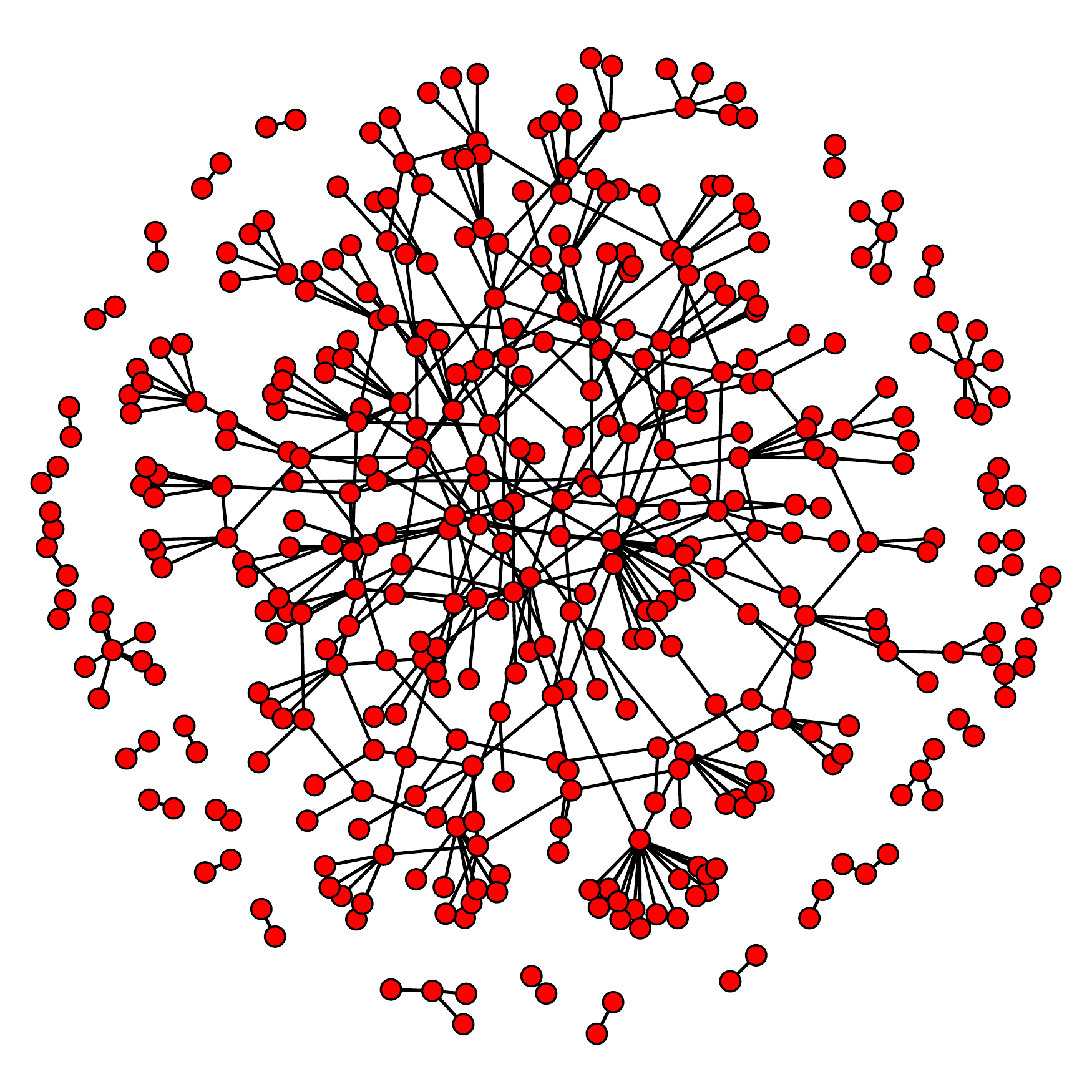} 
\\Geom. weighted degree & -0.014 (-0.033, 0.007)  & reject &
\includegraphics[width=1.1in,height=1.1in]{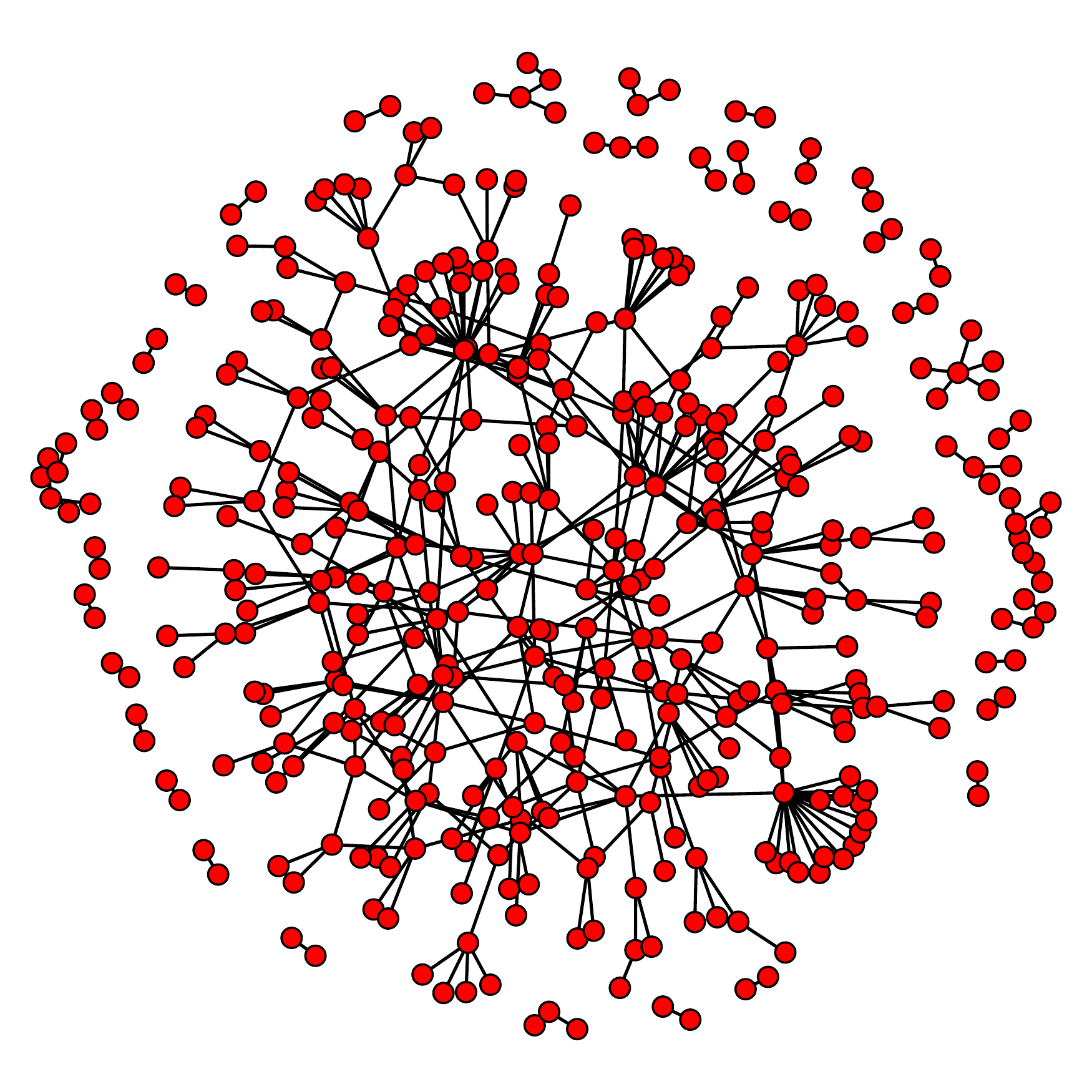} 
\\

\hline
\end{tabular}
\end{table*}



Using the \texttt{ergm} package in R, after fitting the models, we assessed their goodness of fit in the manner described by \cite{hunter2008goodness}, using the \texttt{gof} function with its default settings.  This goodness of fit routine assesses the probability that the distributions of degree, transitive closure and mean geodesic lengths over the nodes in the observed network could have been generated by the fitted model. Results from the \texttt{gof} analysis show that both of the proposed model specifications produce distributions of structural statistics that diverge from the observed values.  Accordingly, the $p-$values for the goodness of fit diagnostics (not shown) indicate rejection of the models.  

Table \ref{ECtab} indicates this and gives values for the SGOF for these models, along with small network visualizations for reference. Although all the models are rejected by structural hypothesis tests, there are marked differences in how well these models fit.  Specifically, the "curved exponential family" version of the model \citep[for more detail, see][]{hunter2006inference} provides a much better fit to the data than the other model without the curved exponential family specification. In fact, at -0.014, the SGOF of this model indicates that it is no better than the null model as an overall description of the structure of the observed data. 
%

The simple lesson here is that goodness of fit based on structural statistics cannot quantitatively distinguish between similar models when all of the models are either accepted or rejected.  Visual inspection of the graphical output can often help in this regard, but is not hard to come up with examples where it cannot.  In these cases it would be good to have an absolute or relative measure of fit to provide a means of model choice.  The AIC is thus a more comparable measure of goodness of fit to the SGOF we propose here, and the following examples make the comparison explicit.

\subsection{Comparison with AIC: Star graph}


The next example considers a 100-node star graph constructed by hand to serve as an imaginary observed network. In addition to the network ties, there is an observed attribute, indicated by the color of the nodes in the visualization.  The attribute values have been measured by our imaginary researcher, but they were not part of the process that generated the network ties.  For this example, we compare the SGOF to AIC from fitted models in the \texttt{ergm} package (Table \ref{startab}).

\begin{table*}[htbp]
\caption{Comparison of Spectral Goodness of Fit to AIC for a star graph}
\label{startab}
  \begin{tabular}{m{1.5in} r r m{1.2in}}
  \hline\hline
Observed Network& & &
\includegraphics[width=1.1in,height=1.1in]{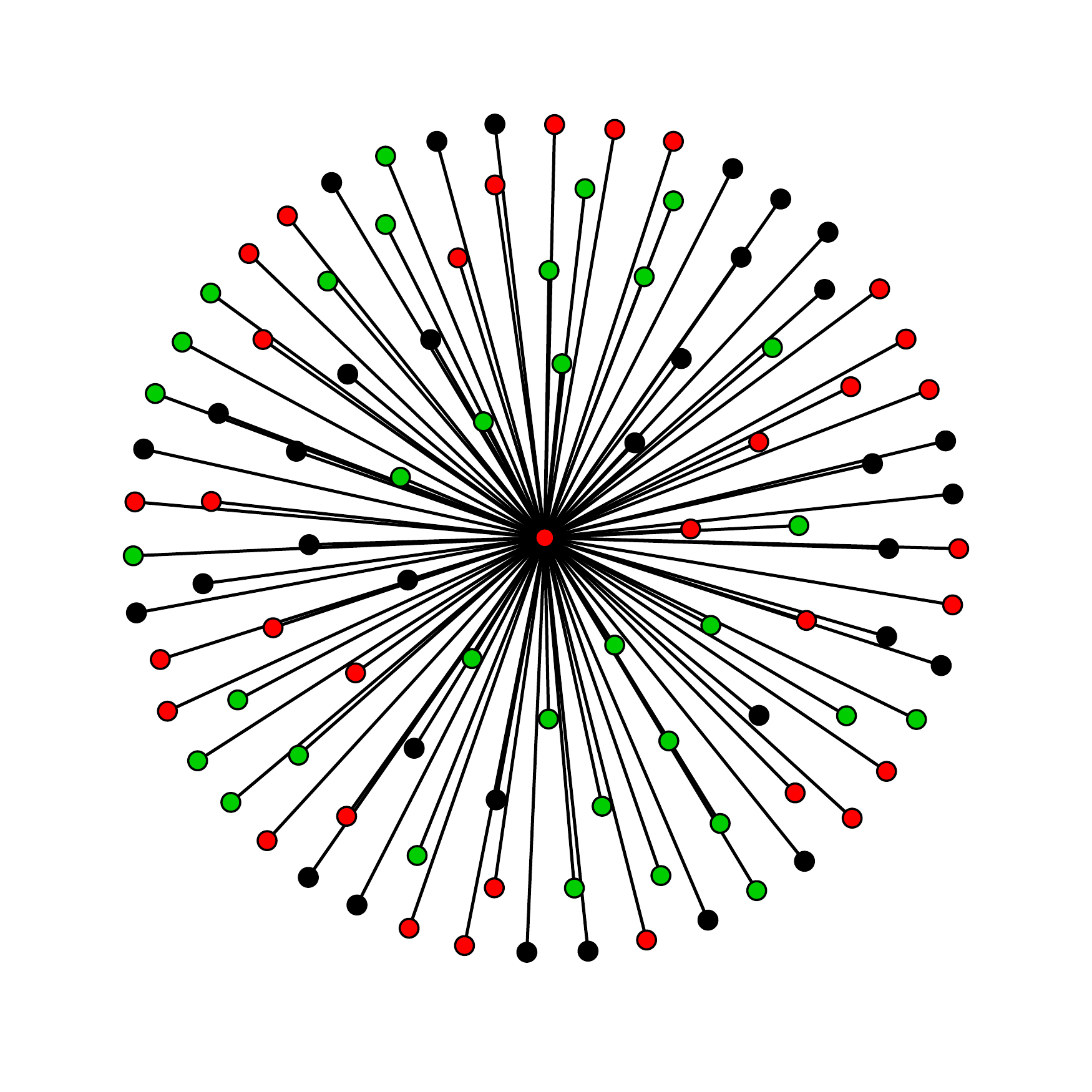} 
\\\hline\\     & SGOF     & Struc. h-test  &  Simulated Network\\\cline{2-4}Null model& 0 (-0.01, 0.014) & 972.59 &
\includegraphics[width=1.1in,height=1.1in]{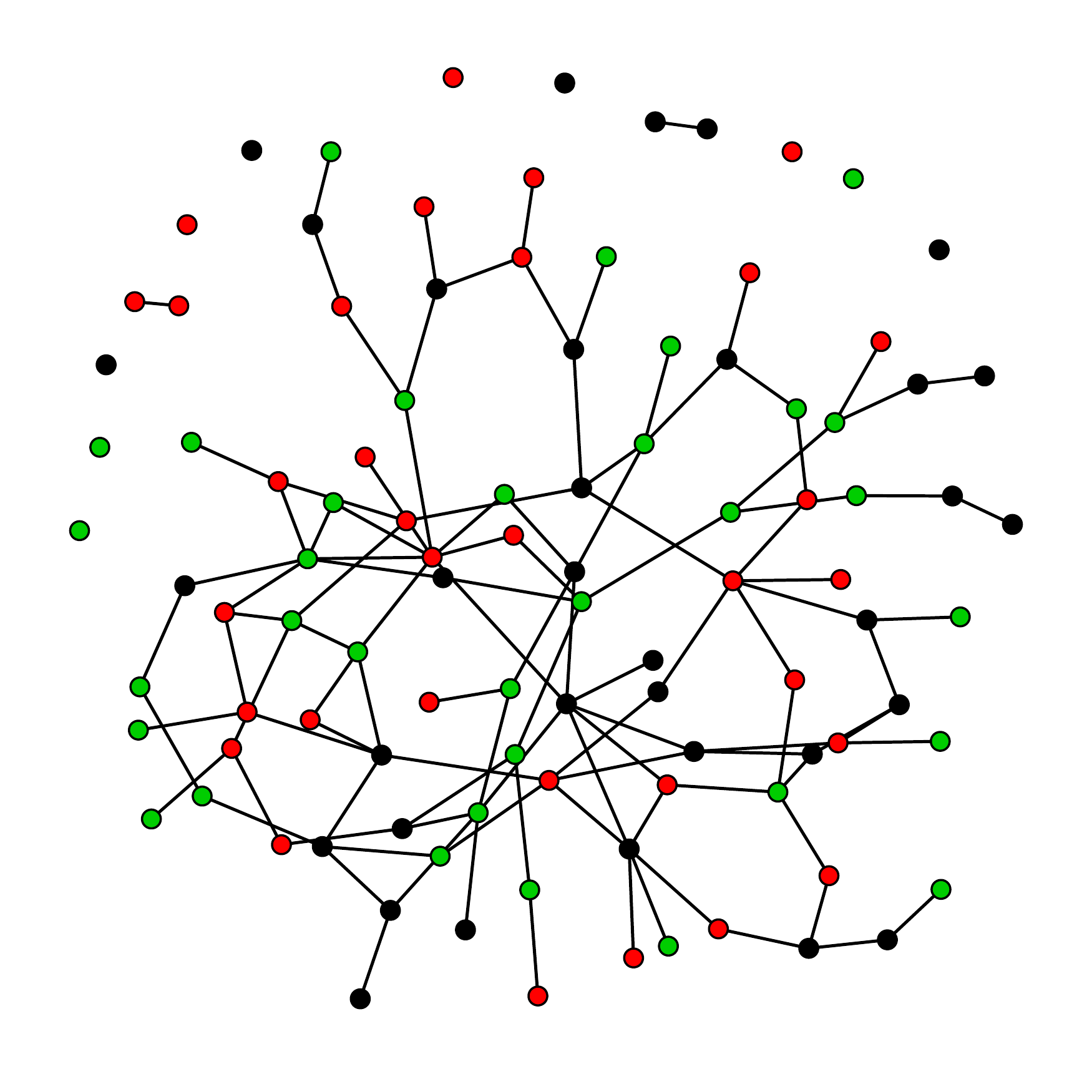} 
\\Red node Homophily& 0.007 (-0.005, 0.025) & 939.83 &
\includegraphics[width=1.1in,height=1.1in]{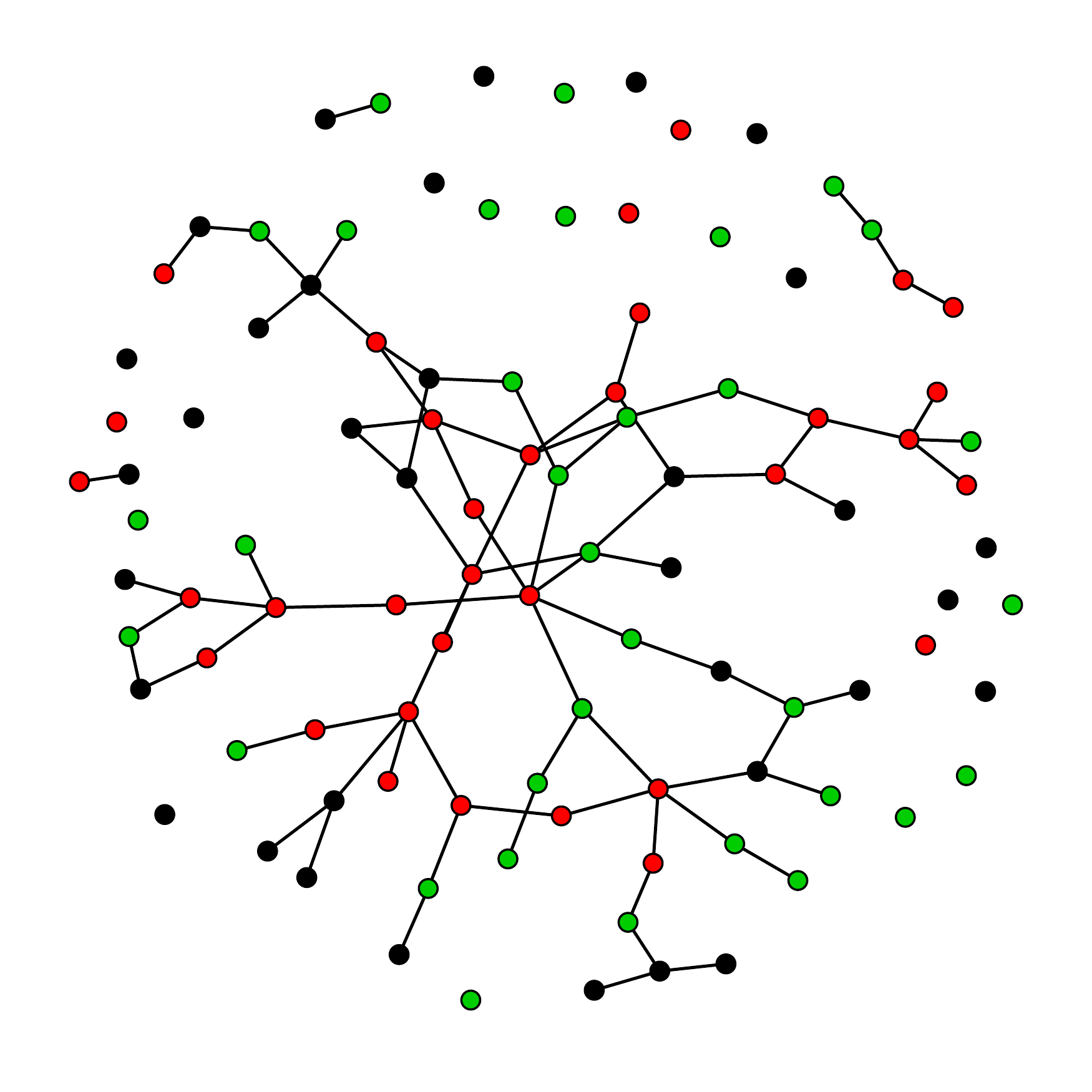} 
\\99-star tendency & 1 (1, 1) & 2322.63 &
\includegraphics[width=1.1in,height=1.1in]{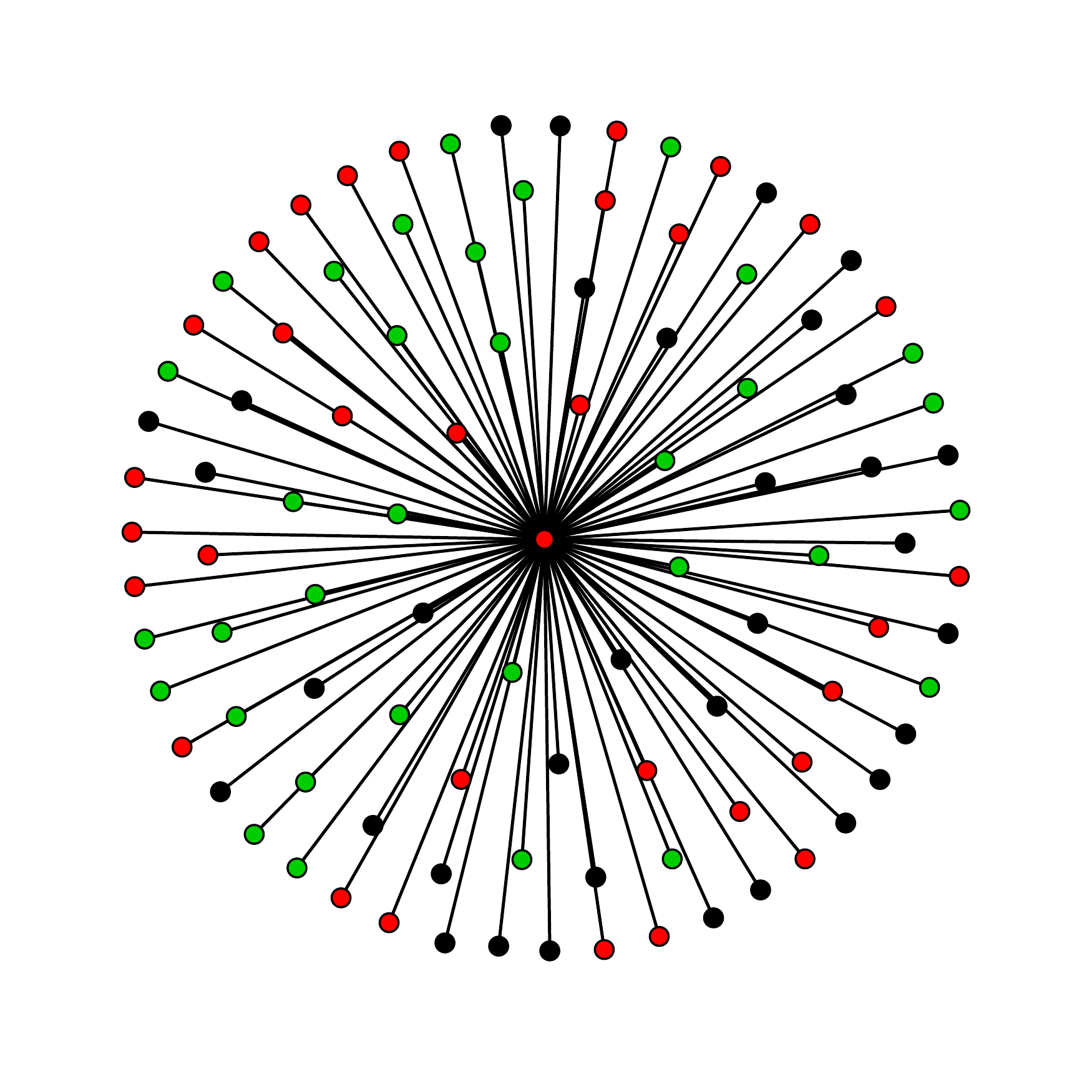} 
\\2-star tendency & 1 (1, 1) & 708.97 &
\includegraphics[width=1.1in,height=1.1in]{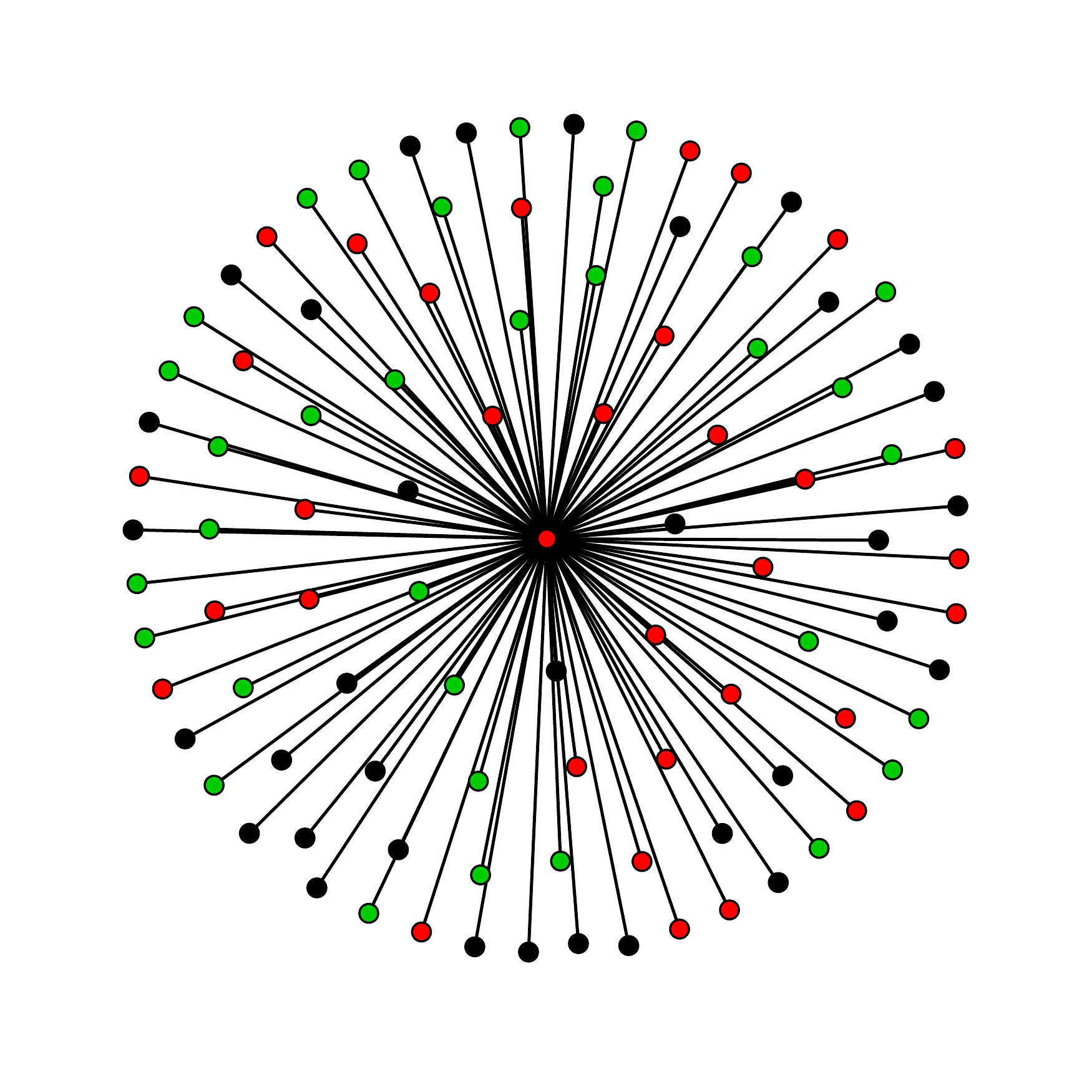} 
\\

\hline
  \end{tabular}
\end{table*}


After the null model, the next model is one fitted with a term for homophily among red nodes in addition to the density term.  The visualization shows that such a model produces a pattern of ties that is very similar to the null model, but a greater proportion of ties among red nodes, similar to the observed network.  It is here that one major difference between SGOF and AIC can be seen.  The SGOF indicates negligible improvement over the null model because the pattern of ties is only a negligible improvement over the null model.  Meanwhile the AIC shows a substantial improvement, from 972.59 to 939.83, because the parameters of the fitted model, including a (spurious by construction) homophily effect, have a higher likelihood than the parameters of the null model, even after accounting for the number of parameters with Akaike's formula. The AIC is senstive to how well the model's parameters fit the data as a whole, including non-structural data.  

The third and fourth models are both ERGMs fit to the data with a $k$-star parameter (tendency toward nodes with degree $k$) in addition to the density parameter, but they differ in how the $k$-star parameter is specified.  The first of the two parameterizes the network with a tendency toward 99-stars, while the second of the two parameterizes the network with a tendency toward two-stars.  Note that the $k$-stars are induced subgraphs, so although there are no nodes with degree two, there are ${99 \choose 2}= 4851$ two-stars, each centered on the same node, while there is only one 99-star in the observed network.  Both of these models produce simulated networks that are star graphs just like the observed network.  Accordingly, the SGOF for both of these models is 1: a perfect fit.  According to the AIC, however, the two models are dramatically different:  the 99-star model is much worse than the null model, with an AIC of 2322.63, while the 2-star model is clearly the best fit of all, with an AIC of 708.97.  Unlike the SGOF, the AIC cannot indicate whether any given fit is good in an absolute sense. 

In practice the AIC and the SGOF are complementary in that they provide answers to different modeling questions.  A researcher may wish to know how well a model fits in terms of both structural effects and nodal or dyadic covariates, or on the other hand, assess the parsimony of the model.  In these cases, the AIC is required.  On the other hand, the researcher may wish to know how well a model that includes both structural effects and nodal and dyadic covariates explains the observed structure, or assess the absolute goodness of fit of a model of structure.  In these cases the SGOF is required.

\subsection{Second comparison to AIC: Faux Mesa High}
The previous example of a star graph was artificially constructed to illustrate the differences between AIC and SGOF.  In this subsection, we give an example of a more typical social network using the "Faux Mesa High" data set of \cite{hunter2008goodness}, adapted from the Add Health surveys \citep{harris2008national}. Similar to the star-graph example, above, after the null model we fit an ERGM model using only homophily effects on the observed covariates, which describe Race, Sex and Grade of the respondents.  We go on to fit a model using only the "Geometrically Weighted Degree" (GWD) of \cite{hunter2006inference} (which is a flexible approach to modeling degree distributions), followed by a model with both the GWD and homophily effects.  The final model differs in type: we consider the preferential attachment model of \cite{barabasi1999emergence}.  Visualizations of the networks created by these models, as well as their AIC and SGOF statistics are shown in Table~\ref{FMHtab}.




\begin{table*}[htbp]
\caption{Comparison of Spectral Goodness of Fit to AIC for Faux Mesa High}
\label{FMHtab}
  \begin{tabular}{m{1.5in} r r m{1.2in}}
  \hline\hline

Observed Network& & &
\includegraphics[width=1.1in,height=1.1in]{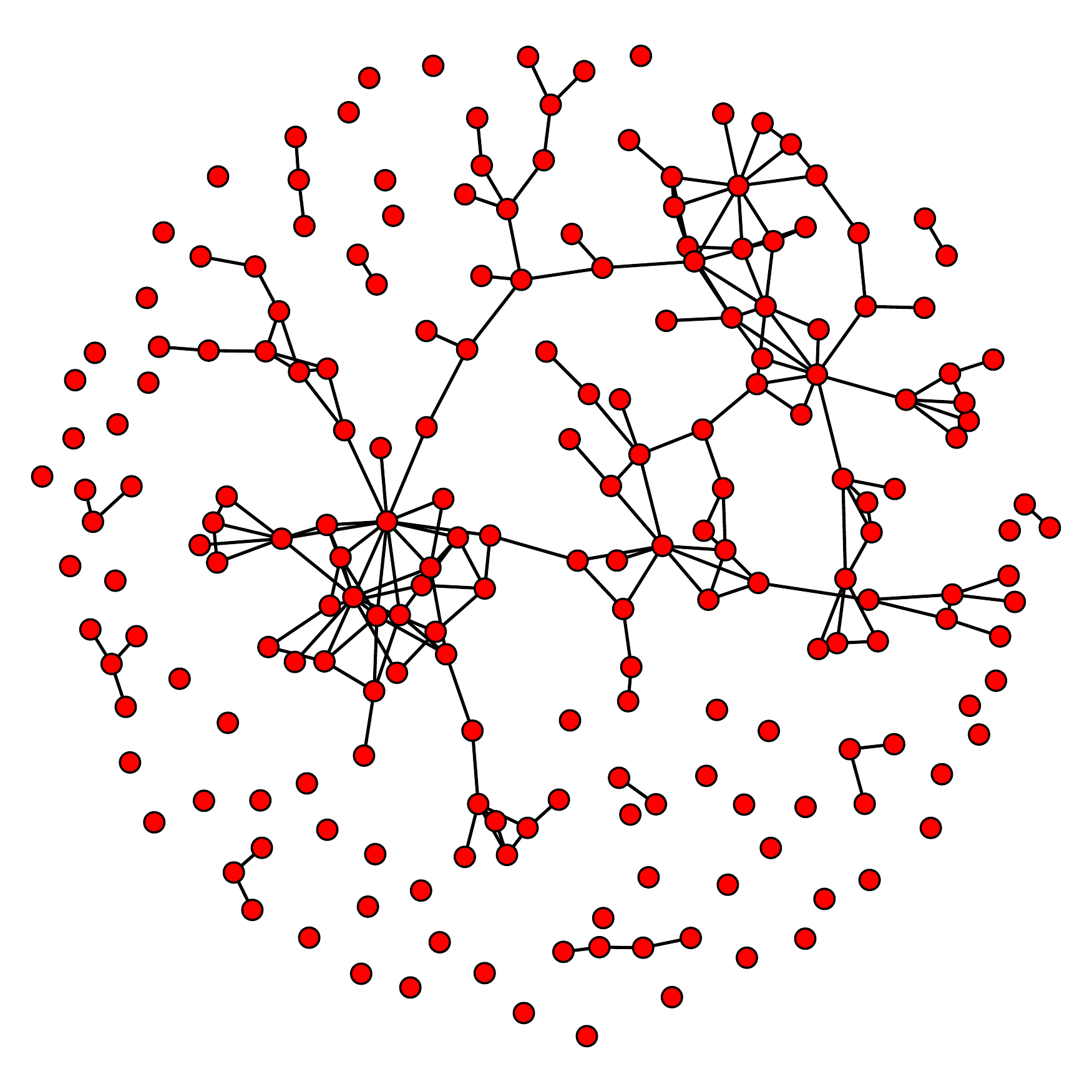} 
\\\hline\\     & SGOF     & Struc. h-test  &  Simulated Network\\\cline{2-4}Null Model & 0 (-0.196, 0.21) & 2287.742 &
\includegraphics[width=1.1in,height=1.1in]{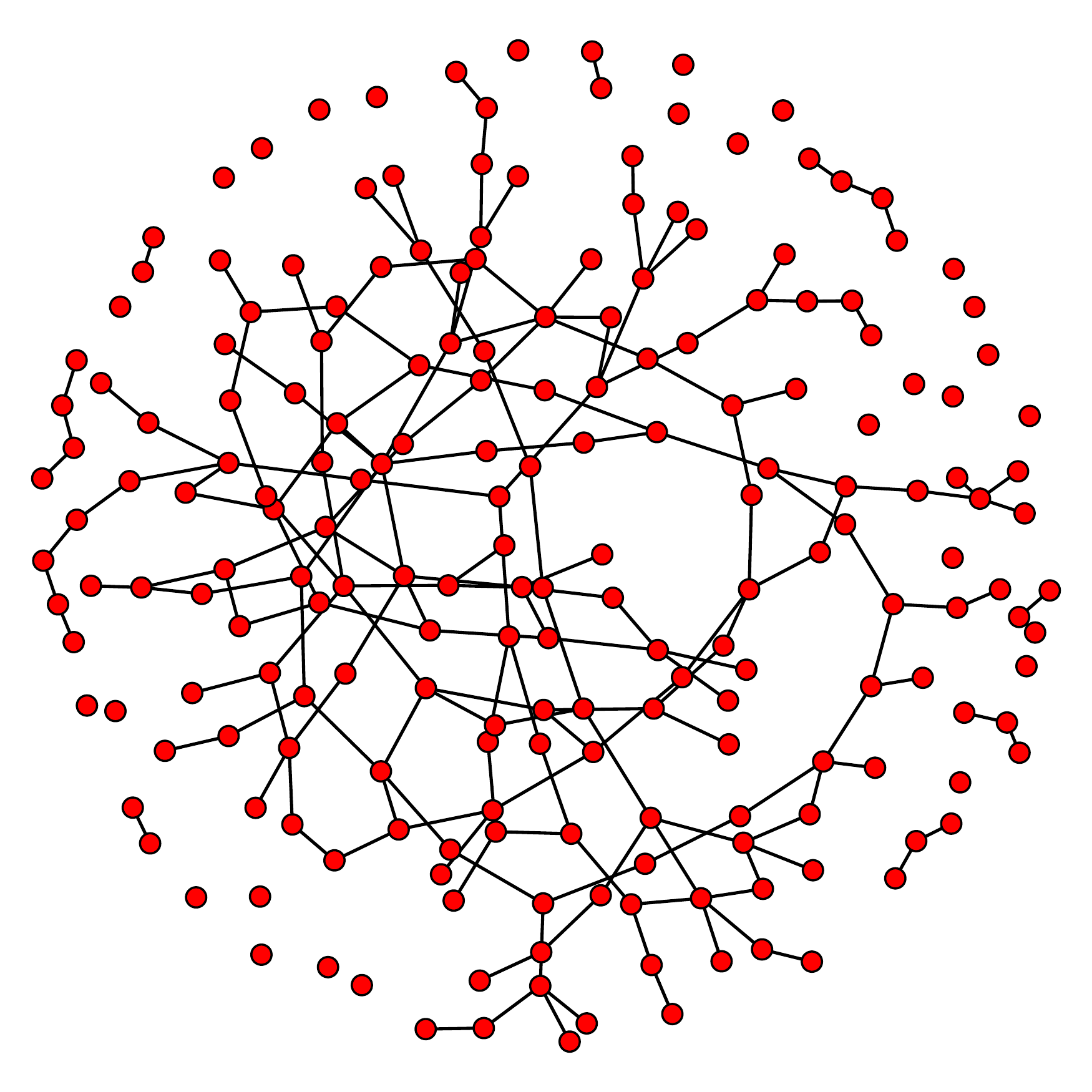} 
\\Homophily on Race, Sex, Grade Only & 0.221 (-0.002, 0.474) & 1890.922 &
\includegraphics[width=1.1in,height=1.1in]{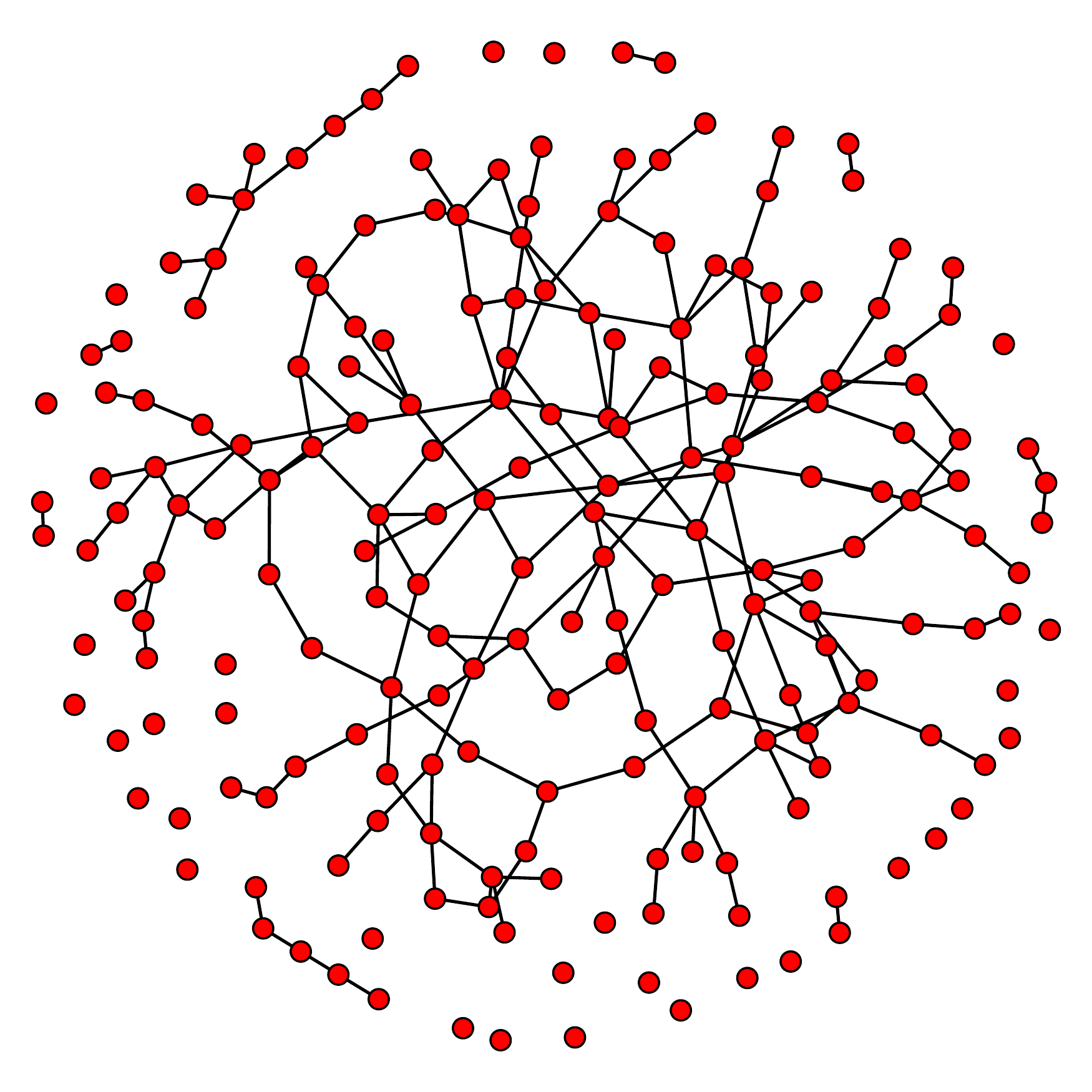} 
\\GWD Only & 0.268 (-0.045, 0.545) & 2245.181 &
\includegraphics[width=1.1in,height=1.1in]{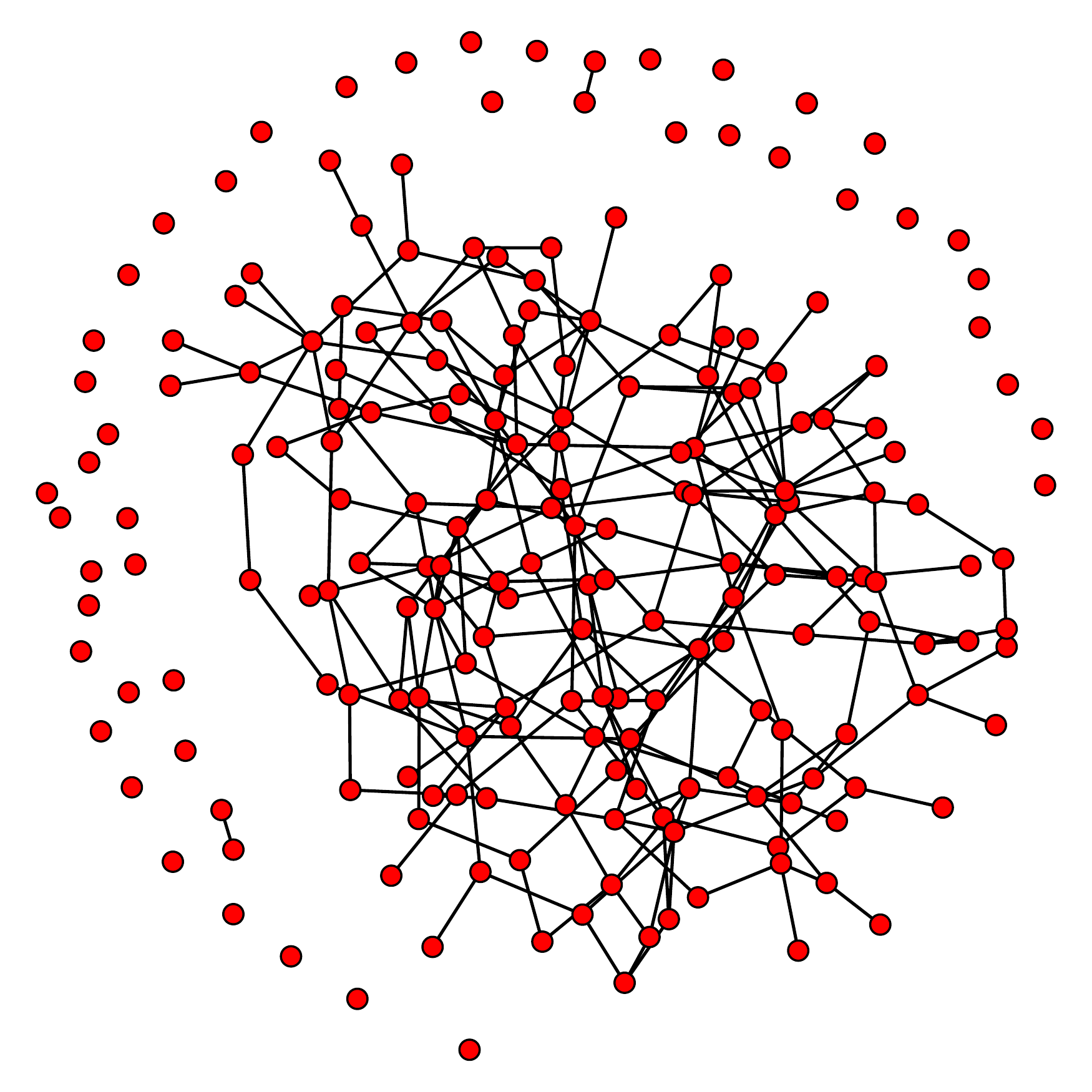} 
\\GWD and homophily & 0.501 (0.259, 0.682) & 1853.656 &
\includegraphics[width=1.1in,height=1.1in]{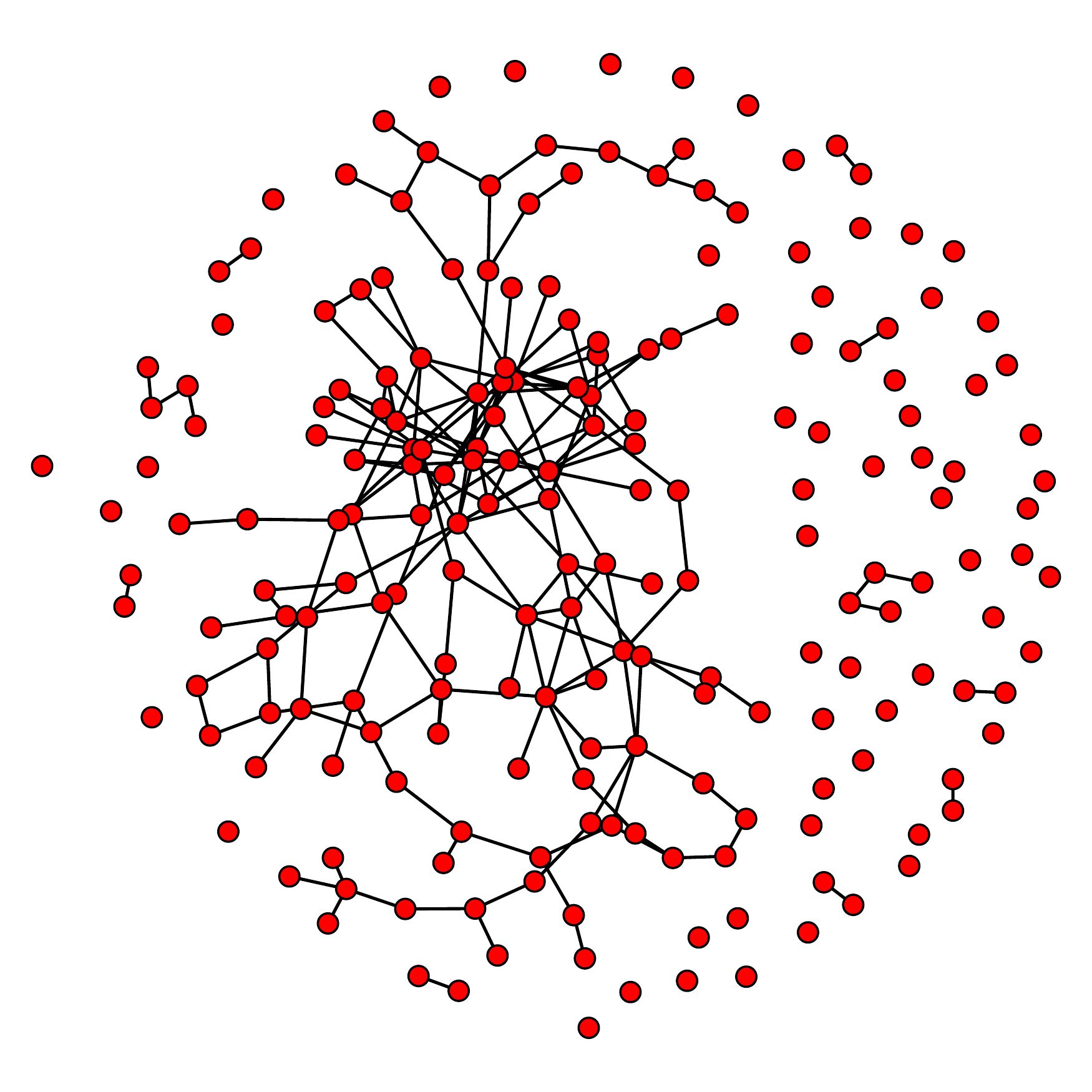} 
\\Preferential attachment& 0.467 (0.16, 0.666) & undefined &
\includegraphics[width=1.1in,height=1.1in]{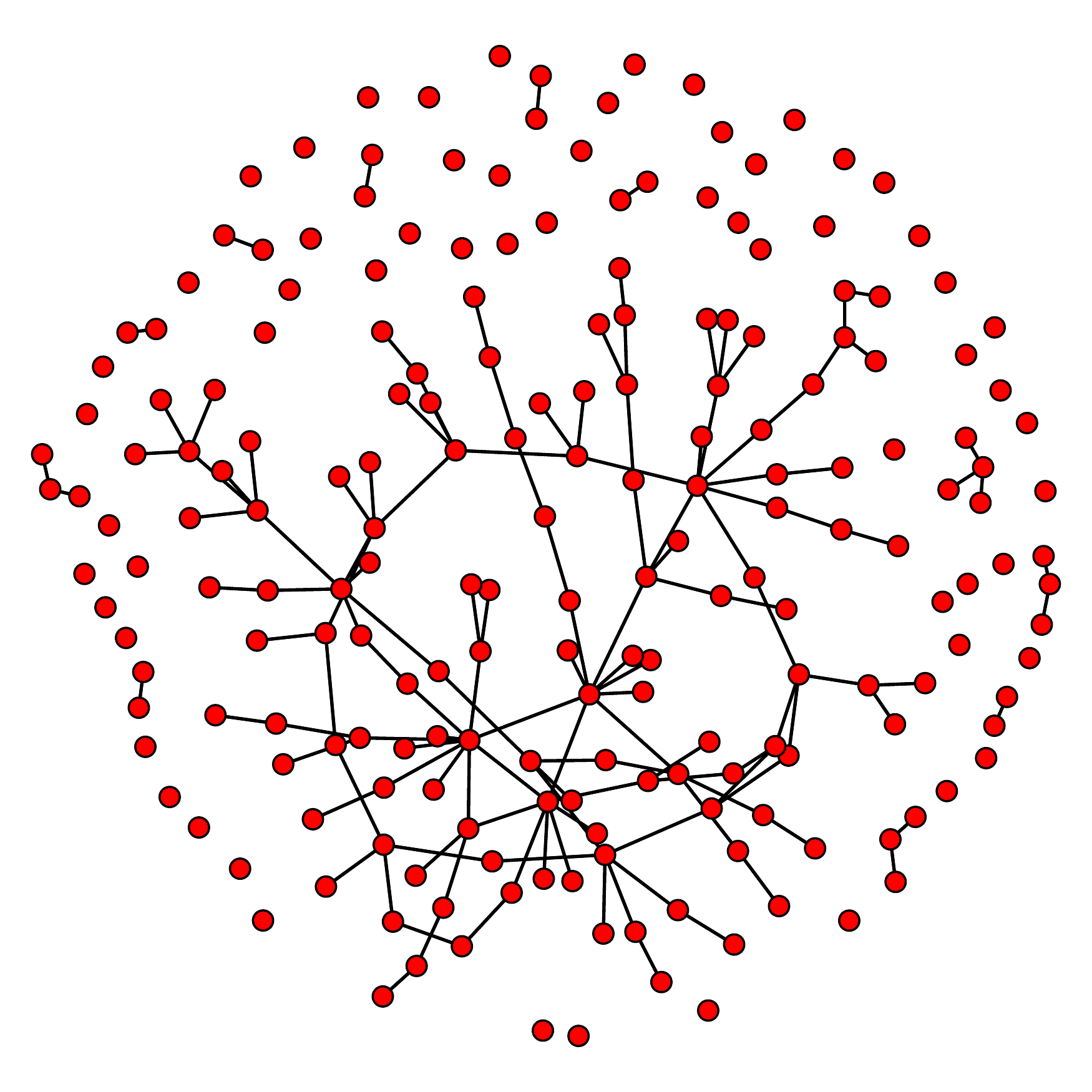} 
\\

\hline
\end{tabular}
\end{table*}

In this example, the homophily on the three covariates makes significant improvements in both SGOF and AIC, because unlike the star graph, there is almost certainly a real homophily effect in the original data.    Likewise, both SGOF and AIC indicate that the model with both GWD and homophily is superior to the models with just one of those two types of effects.  The lessons from Faux Mesa High are, however, otherwise consistent with those from the star graph.  AIC indicates that the homophily-only model is superior to the GWD-only model.  However, from the point of view of generating a pattern of ties alone, the SGOF indicates that the GWD-only model is superior to the homophily-only model. Again, the AIC measures the relative quality of fit of the model as a whole to the data as a whole, while the SGOF measures the absolute quality of the fit of the model to the structure manifest in the observed network ties.  


Finally, we consider a model outside of not only the exponential random graph family, but indeed a model that is algorithmic in nature rather than statistical: the Barab{\'a}si-Albert preferential attachment model \citep{barabasi1999emergence}, as implemented in the \texttt{igraph} package \citep{igraph}. As we use it here, there is no likelihood function and thus no AIC associated with this last model.  The preferential attachment model is based on a generative algorithm with fixed parameters and does not have a likelihood function that could be meaningfully compared to those from fitted ERGMs. 

%


The SGOF is defined, however, as it is for \emph{any model} that generates networks with the same number of nodes as the observed network, regardless of conditions put on the sample space or how (or whether) the model was estimated.  As such, the SGOF makes it possible to compare models that cannot be compared on the basis of the AIC or other likelihood-based methods.  
\subsubsection{Visualization of SGOF}
\begin{figure}[htbp]
\begin{knitrout}
\definecolor{shadecolor}{rgb}{0.969, 0.969, 0.969}\color{fgcolor}
\includegraphics[width=\linewidth]{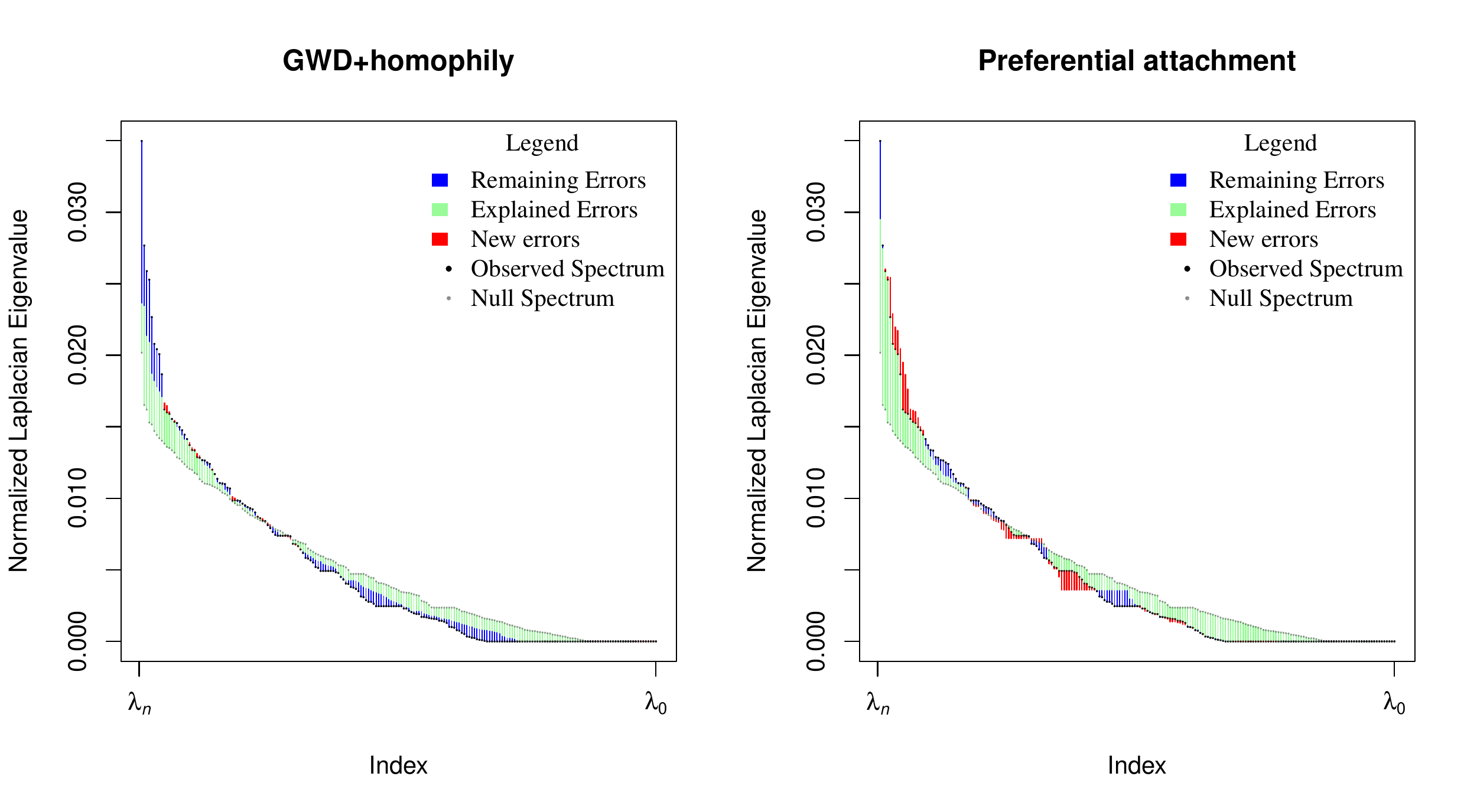} 

\end{knitrout}

\caption{Illustration of spectral qualities of the two best fitted models in Table \ref{FMHtab}.   The green and red indicate improvements and worsening of model fit, respectively, from a change from a null to the fitted model. Blue indicates error left unexplained from the null model.}
\label{errorfig}
\end{figure}

As with other statistical methods, a fuller qualitative understanding of the SGOF can be gained through visualization. Figure \ref{errorfig} plots spectral fits for the ``GWD and Homophily'' and the ``Preferential attachment''  models from Table \ref{FMHtab}, using the \texttt{plotSGOFerrors} function in the \texttt{spectralGOF} package.  

Each panel of the figure is a visualization of spectral error based on three spectra: the observed spectrum, the null model spectrum that is closest to the mean Euclidean distance from the observed spectrum, and the fitted model spectrum that is closest to the mean Euclidean distance from the observed spectrum. The first and the second are the same in both panels and are plotted as points.


The fitted model spectrum is not plotted in points, but rather indicated by colored bars as follows. When the fitted model's spectrum lies between the null and the observed spectra, the fitted model has improved the fit.  The distance between the null and the fitted spectrum is error that has been "explained" and is indicated in light green.  The error that still remains (error that is present under the null and the fitted models) is indicated in blue.

There are also parts of the plots where the fitted and null spectra are on opposite sides of the observed spectrum.  In these cases, the fitted model has "explained" the error between the null and the observed, but introduced new error on the other side of the observed spectrum.  This new error is indicated in red.

Turning to the specific models in Figure \ref{errorfig}, we see that the two fits differ considerably.  In general, the spectrum of the fitted ERGM (left) lies between the observed spectrum and the null spectrum, indicating that the observed network is more structured (farther from random) than are networks simulated from the fitted ERGM.  In contrast, portions of the spectrum of the preferential attachment model (right) are more distant from the null spectrum than is the observed spectrum.  The preferential attachment model has explained more error than the ERGM (represented by more green area in its visualization), but it has also introduced structure not present in the observed network, producing more new error (more red area in the visualization), and resulting in a lower net SGOF.

\subsection{SGOF as an objective function: Collaborations among jazz musicians}

There are sometimes cases when one wishes to implement algorithmic models that do not have an intrinsic means of fitting to observed data.  In this case, SGOF can be useful as an objective function in an exploration of the algorithm's parameter space.  To illustrate this type of application, we consider the network of jazz collaborations described by \cite{gleiser2003community}. 

One theoretically plausible algorithmic model of how collaboration networks are formed is that of \cite{saramaki2004scale}. In this model, one assumes some network exists at $t_0$ to initialize the model.  In subsequent time points, new individuals arrive and form ties to those already present by means of short random walks from a randomly selected node serving as the point of entry into the network. 

 For musicians, the idea would be that after collaborating with some initial partner, one is likely to get to know one's partner's partners, and so on. In addition to being theoretically plausible, this algorithm generates networks with skewed degree distributions and local clustering, as we observe in the jazz collaborations data set.  

To assess the fit of this model, one must first find the best values for the model's parameters, which we will do by appeal to SGOF.  In implementing the algorithm, we left two key parameters to be fitted.  The first is the mean number of edges to add with each new node added to the network.  The second is how many steps in a random walk a new node would take before forming new relationships to existing members of the network. We then generated 100 simulated networks using each combination of parameters, and calculated the SGOF for each pair of parameter values.  

\begin{figure}[htb]
\begin{knitrout}
\definecolor{shadecolor}{rgb}{0.969, 0.969, 0.969}\color{fgcolor}
\includegraphics[width=\linewidth]{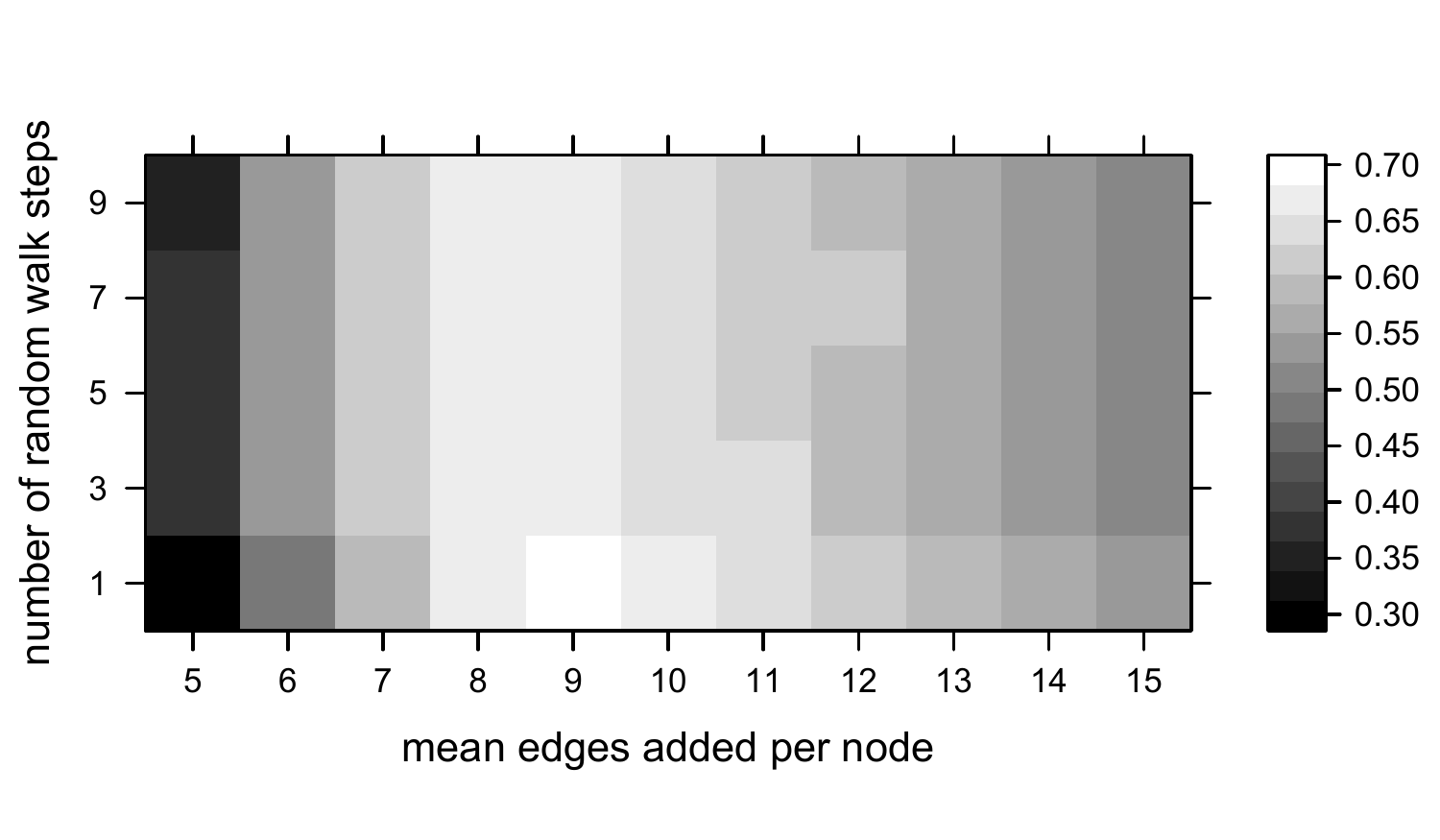} 

\end{knitrout}

\caption{SGOF for different combinations of parameter values for an algorithm based on \cite{saramaki2004scale} fitted to the network of Jazz collaborations described in \cite{gleiser2003community} }
\label{gofGradient}
\end{figure}

The result of this process are shown in Figure \ref{gofGradient}, and indicate that the best fit occurs when the average number of edges added per node is 9, and the random walk distance is a single step.  Thus we can not only use SGOF as a diagnostic tool, but also as a means for identifying the parametric model settings that will be optimal under this criterion.

\section{Future Extensions}
\subsection{Hypothesis testing}
We have presented SGOF as a goodness of fit statistic, analogous to $R^2$. Using spectral distances, it is also possible to construct one and two-sample hypothesis tests for the purposes of formal rejection of certain models in favor of others.  Space does not permit a full discussion of how such tests would be constructed; however, the authors will present this material in a separate manuscript.  

\subsection{Directed graphs}

While the properties of the Laplacian spectrum of undirected graphs have been widely studied and applied, the spectral properties of directed graphs are less well-established. The present paper has therefore focused on undirected, possibly weighted, networks to establish the SGOF, but further work should consider the different properties of directed graphs. For now, we limit ourselves to the following remarks.

The Laplacian matrix for directed networks has been defined differently from that of undirected networks.  In particular, \cite{chung2005laplacians} defines the Laplacian of directed networks as follows.  First, given adjacency matrix, $A$, calculate a matrix, $P$, such that
\begin{equation}
P(i,j)=\frac{A_{ij}}{\displaystyle\sum_k A_{ik}}.
\end{equation}
Then, treating $P$ as the transition matrix of a Markov chain, calculate the Perron vector, $\phi$, which is the all-positive left eigenvector of $P$ corresponding to the stationary distribution of the Markov chain (for strongly connected graphs).  Define $\Phi$ as the matrix with $\phi$ on the diagonal and zeros elsewhere, and $I$ in the standard way as the identity matrix. Finally, the Laplacian for directed graphs is defined as

\begin{equation}
L = I-\frac{\Phi^{1/2}P\Phi^{-1/2}\Phi^{-1/2}P^T\Phi^{1/2}}{2} .
\end{equation}

One feature of this definition is that $L$ is undirected and therefore has real-valued eigenvalues.  Future work should consider the properties of this matrix from the point of view of goodness of fit, but also consider alternative transformations of the adjacency matrix for spectral analysis. 
%
%

\subsection{Statistical properties of Laplacian eigenvalues}

Under certain density conditions, the distribution of eigenvalues of the null model follows the 'semi-circle law' \citep{wigner1955characteristic,chung2003spectra}, but these conditions are restrictive enough that we have chosen to calculate the null errors in the SGOF by simulation rather than by reference to the semi-circle law.  

The statistical properties (e.g. consistency and efficiency) of the eigenvalues of ensembles of networks other than the null model depend on the details of the model from which they are generated, and it is not clear a priori what can be said about the statistical properties  of the SGOF for fitted models in general. As with the null model, the distribution of eigenvalues from certain narrowly defined models have been studied \citep{farkas2001spectra, bolla2004distribution,zhang2014spectra}.  It is not yet clear from the present body of research, however, what can be said about the statistical properties of the SGOF in the general case.

Since we cannot derive the statistical properties of the SGOF analytically, in order to provide one practical point of reference, we have conducted a simulation-based exploration of the properties of 100-node density-only models, under a range of densities.  These simulations support the following tentative conclusions. The means of individual eigenvalues are stable across sample sizes (where sample size refers to the number of simulated networks from which the mean spectrum is calculated).  The standard deviations of individual eigenvalues from Erd\H{o}s-R\'enyi random graphs are asymptotically consistent, but biased downwards for small numbers of simulated networks. Likewise, the $5^{th}$ and $95^{th}$ quantiles of individual eigenvalues are asymptotically consistent, but biased toward the median for small samples of simulated networks.

Given the above,  we recommend using 100 simulations of the null model to calculate standard errors or quantiles of the distribution of SGOF for exploratory modeling and at least 1000 simulations for published results. Furthermore, we strongly recommend examining the distribution of spectra simulated from fitted models to establish that sufficient sample sizes have been obtained when calculating the SGOF.  Future work should seek to derive more general conclusions about the statistical properties of spectral distances for network models.

\section{Conclusion}

We have proposed a new measure of goodness of fit for network models based on the spectrum of the graph Laplacian: "spectral goodness of fit" (SGOF), and provided code with which SGOF can be easily implemented.  The properties of SGOF fill gaps left by the current set of goodness of fit indicators, making it complementary to existing methods.  

Table \ref{summarytab} summarizes the properties of each approach to goodness of fit.  Analogous to the standard $R^2$, the SGOF statistic measures the percent improvement in network structure explained over a null model.  By measuring fit relative to fixed reference points, SGOF can be said to provide an "absolute" measure of goodness of fit.

\begin{table}[htb]
\caption{Summary of properties of goodness of fit measures}
\label{summarytab}
  \begin{tabular}{l ccc }
  \hline\hline
                                    &Struct. stats&AIC&SGOF\\
                                    \cline{2-4}
Absolute Measure of GOF&&&Yes\\
Relative Measure of GOF&&Yes&Yes\\
Sensitive to structure only&Yes&&Yes\\
Hypothesis test of model fit&Yes&&\\
Sensitive to model specification&&Yes&\\
Requires Likelihood Function&&Yes&\\

\hline
\end{tabular}

\end{table}

Prior methods had provided relative measures of fit (AIC), and hypothesis testing of fit for specific subgraph statistics, but until now there was no absolute measure of fit for network structure as a whole. Ultimately, however, we see SGOF as playing a complementary role to existing techniques.  For example, when a research question concerns a specific structural tendency (say, to transitive closure), one should use both structural statistics as well as SGOF (and even AIC if applicable, to assess model parsimony).

In addition to providing an absolute measure of fit, the SGOF allows the comparison of models fit by diverse means and of diverse functional forms.  We hope that the ability to compare fit among dissimilar models will facilitate building on and refining prior work, as well as greater engagement with research models and results from outside of any given researcher's own methodological tradition.


\section*{References Cited}

\bibliographystyle{plainnat} 
\bibliography{spectralGOF}

\end{document}